\documentclass{iaus}
\usepackage{graphicx}
\usepackage{natbib}
\usepackage{aas_macros}

\title[Investigation of a Sunspot Complex
by  Helioseismology] 
{Investigation of a Sunspot Complex by Time-Distance Helioseismology}

\author[A.~G.~Kosovichev \& T.~L.~Duvall~Jr]   
{A.~G.~Kosovichev$^1$
 \and T.~.L.~Duvall~Jr$^2$}
\affiliation{$^1$Stanford University, Stanford, CA 94305, USA \\
$^2$Solar Physics Laboratory, Goddard Space Fight Center, NASA,
Greenbelt, MD 20771, USA
}

\pubyear{2010}
\volume{273}  
\pagerange{119--126}
\jname{Physics of Sun and Star Spots}
\begin{document}

\maketitle

\begin{abstract}
Sunspot regions often form complexes of activity that may live for
several solar rotations, and represent a major component of the
Sun's magnetic activity. It had been suggested that the close
appearance of active regions in space and time might be related to
common subsurface roots, or "nests" of activity. EUV images show that
the active regions are magnetically connected in the corona, but
subsurface connections have not been established. We investigate the
subsurface structure and dynamics of a large complex of activity,
NOAA 10987-10989, observed during the SOHO/MDI Dynamics run in
March-April 2008, which was a part of the Whole Heliospheric
Interval (WHI) campaign. The active regions in this complex appeared
in a narrow latitudinal range, probably representing a subsurface
toroidal flux tube. We use the MDI full-disk Dopplergrams to measure
perturbations of travel times of acoustic waves traveling to various
depths by using time-distance
helioseismology, and obtain sound-speed and flow maps by
inversion of the travel times. The subsurface flow maps show an interesting dynamics of
decaying active regions with persistent shearing flows, which may be
important for driving the flaring and CME activity, observed during
the WHI campaign. Our analysis, including the seismic sound-speed
inversion results and the distribution of
deep-focus travel-time anomalies, gave indications of diverging
roots of the magnetic structures, as could be expected from
$\Omega$-loop structures. However, no clear connection in
the depth range of 0-48 Mm among the three active regions in this
complex of activity was detected.

\keywords{Sun: helioseismology, sunspots, Sun: interior, Sun: magnetic field}
\end{abstract}

\firstsection 
\section{Introduction}
Local helioseismology provides important insight
into the subsurface structure and dynamics of emerging magnetic
flux, formation, evolution and decay of sunspots and active regions.
Methods of local helioseismology and acoustic tomography, based
on measurements and inversion of acoustic travel times,
are intensively developed and tested via numerical
simulations \citep[for a recent review see,][]{Kosovichev2010}.
These methods
provide important insights in the subsurface structure and dynamics of sunspots,
which are important for understanding the origins of solar magnetism.
In particular, our previous results obtained by a time-distance helioseismology
techniques \citep{Duvall1993} revealed significant changes in the subsurface flow patterns
during the life cycle of active regions: strong diverging flows
during the flux emergence, formation of localized converging flows
around stable sunspots, and dominant outflows during the decay \citep{Kosovichev2006,Kosovichev2009}.
The sound-speed images indicate that the magnetic flux gets concentrated
in strong field structures just below the surface during the sunspot
formation, but the seismic perturbations of large sunspots extend
at least up to 20-30 Mm, indicating that sunspots have deep roots.
\begin{figure}[t]
 \begin{center}
 \includegraphics[width=0.6\textwidth]{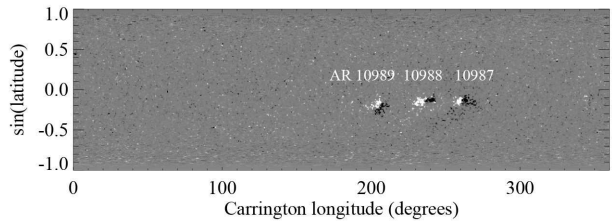}
  \caption{Synoptic magnetic field map from SOHO/MDI for Carrington rotation 2068.
The grayscale shows the line-of-sight magnetic field saturated at $\pm 100$ G.}\label{fig1}
  \end{center}
\end{figure}

A complex of decaying sunspots, NOAA 10987-10989, was observed by
SOHO/MDI in March 2008 during  Whole Heliospheric Interval (WHI)
campaign. Three active regions appeared in a narrow latitudinal and
longitudinal range (Fig.~\ref{fig1}). This  suggests that they
probably originated from a common subsurface nest of activity.

\section{Data Analysis and Results}

For the analysis we used full-disk Dopplergrams from SOHO/MDI
\citep{Scherrer1995} obtained with 1-min cadence and 2 arcsec/pixel
resolution. The time-distance data analysis and inversion procedure
is described by \citet{Duvall1997,Kosovichev1996,Kosovichev1997}.
The acoustic travel-time maps were calculated for 12 distance ranges
(annuli), from 0.78 to 11.76 heliographic degrees (or from 9.48 to 142.85
Mm) for fifteen 8-hour intervals during the period of March 27 - April 1, 2008. We used
both the surface-focus and deep-focus measurement schemes
\cite{Duvall1998}. This  coverage allowed us to investigate
evolution of this decaying complex of activity.
\begin{figure}[t]
 \begin{center}
 \includegraphics[width=0.95\textwidth]{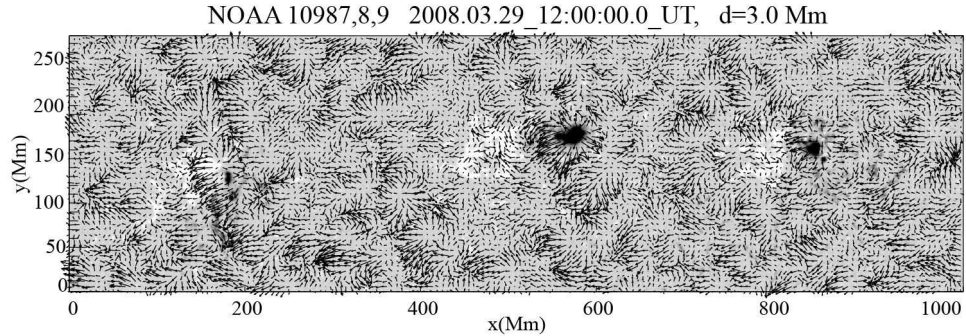}
  \caption{A map of subsurface flows at depth 3Mm (arrows)
  over the corresponding photospheric magnetogram for all three active regions,
  observed on March 29, 2008, 12:00 UT. The longest arrow correspond to 0.5~km/s.
The magnetic field scale is from -100 to 500 G.}\label{fig2}
  \end{center}
\end{figure}

Figure~\ref{fig2} shows a subsurface flow map of all three active
regions at the depth of 3~Mm. The flow pattern shows
supergranulation flows and strong diverging flows around sunspots of
these active regions. The diverging flows are observed in the whole
range of depth, $0-12$~Mm obtained in our inversions. At greater
depths the inversion results become too noisy. We did not attempt to
recover the flow patterns at greater depths in this work. The
dominance of diverging flows for the decaying sunspots is consistent
with our previous results \citep{Kosovichev2006,Kosovichev2009}.
\begin{figure}[t]
 \begin{center}
 \includegraphics[width=0.75\textwidth]{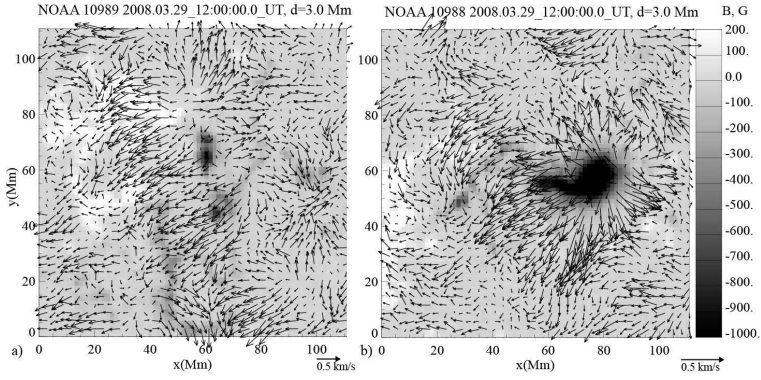}
  \caption{Subsurface flow maps at depth 3 Mm in active regions a) 10989, and b) 10988.
 The background images are the corresponding photospheric magnetograms. }\label{fig3}
  \end{center}
\end{figure}

The structures of the subsurface flows in AR 10989 and 10988 are
shown in more detail in Figure~\ref{fig3}. An interesting feature is
strong shearing flows associated with the diverging flows of the
decaying sunspots. These flows may be important for understanding
the flaring and CME activity of decaying regions. In this case,
during the WHI campaign several strong CMEs were detected from AR
10989, which was decaying most rapidly.
\begin{figure}[t]
 \begin{center}
 \includegraphics[width=0.8\textwidth]{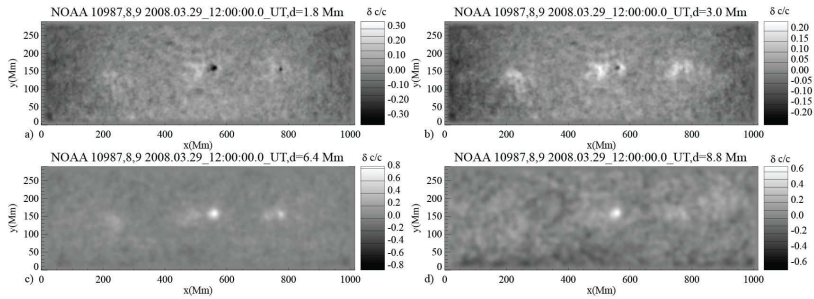}
  \caption{Maps of subsurface sound-speed variations beneath the active regions,
NOAA 10987 -- 9, at depth: a) 1.8 Mm, b) 3 Mm, c) 6.4 Mm, and d)
8.8 Mm.}\label{fig4}
  \end{center}
\end{figure}

Our inspection of the flow maps did not reveal any large-scale
pattern, which can be common for these active regions and indicate
on their subsurface connections. Therefore, we investigated also
helioseismic sound-speed maps, obtained by inversion of the
surface-focus travel-time variations, using the ray-path
approximation \citep{Kosovichev1999}. Samples of the seismic
sound-speed maps are shown in Fig.~\ref{fig4}. Interestingly enough,
a shallow layer of negative sound-speed variation, previously
reported for isolated sunspots
\citep[e.g.][]{Couvidat2006,Couvidat2006a,Kosovichev2000,Zhao2010}, is
observed only for the leading sunspots of AR 10988 and 10987. The
seismic structure of the sunspot in AR 10989 showed only positive
variations, probably because it was mostly decayed.

\begin{figure}[t]
 \begin{center}
 \includegraphics[width=0.8\textwidth]{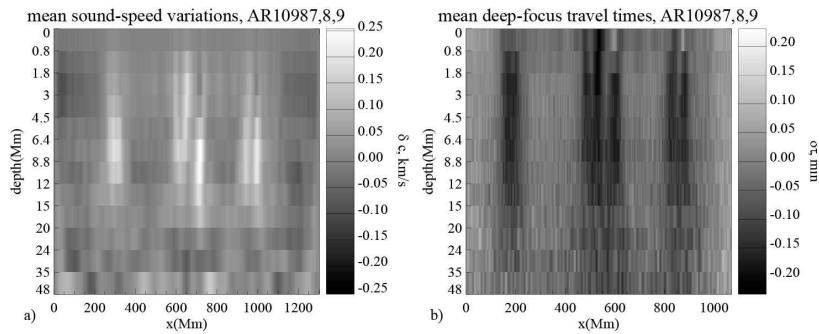}
  \caption{a) Sound-speed variations, averaged over a narrow range of latitude
including all three active regions, as a function of longitude
($x$-coordinate) and depth. b) Travel-time variations calculated by
using the deep-focus scheme and averaged over a narrow range
of latitude, as a function of longitude and the focus depth.}\label{fig5}
  \end{center}
\end{figure}

These maps, in which the seismic structure is clearly seen up to 12
Mm depth, also did not reveal potential subsurface links among the
active regions. Therefore, we studied averaged properties in order
to obtain a better signal-to-noise ratio at greater depths.
Figure~\ref{fig5}$a$  shows the depth structure of the sound-speed
inversions averaged over a narrow range of latitude.
The seismic structures can be traced up to the
depth of 24 Mm. It seems that the tube-like structures become more inclined and diverge
from each other with depth, like this can be expected from $\Omega$-loop structures,
but this evidence is weak. From these inversion data we cannot make
a conclusion about the subsurface connections.

In addition, we attempted to detect the links directly from the
deep-focus travel time anomalies. Figure~\ref{fig5}$b$ shows the
deep-focus travel time perturbations (relatively to the quiet Sun)
averaged over a range of latitudes including the active regions, as
a function of longitude and the focus-point depth.
The depth distribution of these travel-times is qualitatively similar
to the sound-speed inversion results from the different
(surface-focus) type of measurements. The deep-focus data allows us
to trace the subsurface structures in deeper layers, up to 48 Mm.
They also indicate divergence of the active region structures, which
may be an evidence that these are parts of a $\Omega$-loop toroidal
structure. But this remains uncertain and requires further
investigation.

\section{Conclusions}
We have used data from SOHO/MDI to investigate seismic sound-speed
structures and mass flows in the solar interior, associated with a
complex of three active regions. We found that all sunspots in this
complex were decaying and were surrounded by strong deep outflows
and shearing flows, which could be important for initiation of CMEs
observed during this period. Our time-distance helioseismology
analyses, including seismic sound-speed inversions of the
surface-focus travel-time measurements and deep-focus travel-time
anomalies, provided indications of diverging roots of the magnetic
structures as this could be expected from $\Omega$-loop structures, but
were unable to detect connections in the depth range of 0-48 Mm
among the three active regions in this complex.



\bibliographystyle{aa}




\end{document}